\documentclass[conference,letterpaper]{IEEEtran}
\IEEEoverridecommandlockouts
\usepackage[utf8]{inputenc} 
\usepackage[T1]{fontenc}
\usepackage{url}              
\usepackage{ifthen}          
\usepackage{cite}             
\usepackage{tikz}
\usepackage{amsmath, amssymb}  
\usepackage{graphicx}         
\usepackage{booktabs}         

\usepackage{xcolor}
\usepackage{caption}
\newtheorem{thm}{{{\textit{Theorem}}}}

\newtheorem{lemma}{{{\textit{Lemma}}}}

\newtheorem{construction}{{{\textit{Construction}}}}
\newtheorem{defn}{{{\textit{Definition}}}}

\newtheorem{example}{{{\textit{Example}}}}

\newcounter{cases}
\newcounter{subcases}[cases]
\newenvironment{mycase}
{
	\setcounter{cases}{0}
	\setcounter{subcases}{0}
	\newcommand{\case}
	{
		\par\indent\stepcounter{cases}\textit{Case \thecases.}
	}
	\newcommand{\subcase}
	{
		\par\indent\stepcounter{subcases}\textit{Subcase (\thesubcases):}
	}
}
{
	\par
}
\renewcommand*\thecases{\arabic{cases}}
\renewcommand*\thesubcases{\roman{subcases}}
\begin{document}
\title{A New Construction of Optimal Symmetrical ZCCS} 

%
%
 \author{%
   \IEEEauthorblockN{Rajen Kumar}
   
   \IEEEauthorblockA{Department of Mathematics\\
                     Indian Institute of Technology\\ 
                     Patna, India\\
                     rajen\_2021ma04@iitp.ac.in}
\and
 \IEEEauthorblockN{Prashant Kumar Srivastava}
   
   \IEEEauthorblockA{Department of Mathematics\\
                     Indian Institute of Technology\\ 
                     Patna, India\\
                      pksri@iitp.ac.in}
\and
                     \IEEEauthorblockN{Sudhan Majhi}
   \IEEEauthorblockA{Department of ECE\\
    Indian Institute of Science\\
                     Bangalore, India\\
                     smajhi@iisc.ac.in}
                    
  \thanks{The work of Rajen Kumar is supported in part by the CSIR, India, under award letter $09/1023(0034)/2019$-EMR-I.}
  \thanks{The work of Sudhan Majhi
is supported by the SERB, Govt. of India, under grant no. CRG$/2022/000529$ and EEQ$/2022/001018$.}}
%
%

\maketitle

\begin{abstract}
 We propose new constructions for a two-dimensional ($2$D) perfect array, complete complementary code (CCC), and multiple CCCs as an optimal symmetrical $Z$-complementary code set (ZCCS). We propose a method to generate a two-dimensional perfect array and CCC. By utilising mutually orthogonal sequences, we developed a method to extend the length of a CCC without affecting the set or code size. Additionally, this concept is extended to include the development of multiple CCCs, and the correlation characteristics of these multiple CCCs are identical with the characteristics of optimal symmetrical ZCCS.
\end{abstract}
\section{Introduction}
The widespread investigation of high-dimensional signals with impulse-like high-dimensional periodic auto-correlation has been driven by the application of synchronisation\cite{Golomb_TIT_1982,Hershey_EL_1983}, phased-array antennas \cite{Golay_71}, communication\cite{Fan1996SEQUENCE}, radar\cite{Weathers_1983}, sonar, and other fields. Numerous methods have been presented to generate such signals. There are two significant groups of such signals. The first one is known as a high-dimensional perfect array, while the second one is referred to as a high-dimensional complementary array set. The high-dimensional periodic auto-correlation function (PACF) is derived by summing the auto-correlation functions of all sub-arrays inside it. Considering the ease of implementing signals, two-dimensional ($2$D) signals are favoured.

Multi-carrier code division multiple access (MC-CDMA) systems that employ Complete Complementary Codes (CCC) have various advantages compared to traditional CDMA systems. These advantages include offset stacked spreading, being able to provide results in lower power spectral density, multi-access interference-free transmission, more resistance to jamming and offering more freedom in positioning code elements within a channel \cite{Trucs}. In MC-CDMA, each code is allocated to a unique user, resulting in a need for a substantial quantity of codes. The rise in the number of codes in CCC results in an expansion of the code size. CCC with various parameters can be found in \cite{sarkar2024construction, Praveen_CCC, Tao_SPL_ZCCS}. The set size of the $Z$-complementary code set (ZCCS) is considerably larger than that of CCC, which promotes the ability of a ZCCS based MC-CDMA system to support a larger number of users, especially in a quasi-synchronous environment \cite{Fan}.

The idea of a type-I ZCCS is proposed to overcome this drawback by reducing the dependence on ideal correlation from all delays to a zero correlation zone (ZCZ) in the front end, which results in the set size of type-I ZCCS being several times higher than that of CCCs \cite{Feng_ZCCS}. Recently, Kumar \textit{et~al.} proposed a construction for type-II ZCCS with ZCZ at the tail end, providing more codes than the type-I ZCCS with the same flock size \cite{Rajen}. Another type, symmetrical ZCCS, was introduced by Zhau \textit{et~al.} with ZCZ at both the front and tail end \cite{Zhou_sym_ZCCS_2023}. Multiple CCCs (MCCC) with favourable inter-set cross-correlation attributes have received recent attention from researchers for their ability to resist inter-cell interference caused by multiple cells in multi-cell MC-CDMA systems \cite{Tian_multiple_CCC}. The primary objective of this work is to explore construction techniques that incorporate flexible factors. We have made devoted attempts to fulfil this objective.

The remaining portion of the work is structured as follows. Section \ref{Sec:Pre} presents preliminary information on the notations and definitions. In Section \ref{Sec:PC}, we presented proposed constructions combined with a complete explanation and an illustrative example. Section \ref{Sec:comparision} provides a comparison to previous research. Finally, in Section \ref{sec:conclusion}, a conclusion has been made.

\section{Preliminaries}\label{Sec:Pre}
In this section, we provide a few important definitions and symbols that will be utilized throughout the paper.
\begin{defn}
Let 
\begin{equation}\label{Eq:code}
    C_k=\begin{bmatrix}
         c_{1,1}^k & c_{1,2}^k  &\cdots & c_{1,N}^k \\
         c_{2,1}^k  & c_{2,2}^k  &\cdots & c_{2,N}^k \\
         \vdots &\vdots &\vdots &\vdots\\  c_{M,1}^k  & c_{M,2}^k  &\cdots & c_{M,N}^k 
    \end{bmatrix}
\end{equation}
be a code of size $M\times N$. Then $2$D PACF $\mathcal{P}(C_k)(\tau_1,\tau_2)$ is defined as follows
\begin{equation}
    \mathcal{P}(C_k)(\tau_{1}, \tau_{2})=\sum\limits_{i_{1}=1}^{M}\sum\limits_{i_{2}=1}^{N}c_{i_{1},i_{2}}^k(c_{i_{1}+\tau_{1},i_{2}+\tau_{2}}^k)^{*},
\end{equation}
where the symbol ``$*$'' denotes the complex conjugate and shifts $|\tau_1|<M, |\tau_2|<N$ subscript addition $i_1+\tau_1$ and $i_2+\tau_2$ are calculated modulo $M$ and $N$, respectively.
\end{defn}
\begin{defn}
    Let $C$ be a code defined in \eqref{Eq:code} and referred to as a perfect array if its $2$D PACF has
    \begin{equation}
        \mathcal{P}(C)(\tau_{1},\tau_{2})=\left\{ \begin{array}{cc}
            NM & (\tau_{1}, \tau_{2})=(0,0) \\
            0 & (\tau_{1}, \tau_{2})\neq(0,0)\end{array}\right..
    \end{equation}
\end{defn}

 \begin{defn}
     Let $C_{s}$ and $C_{t} $ be any two codes of order $M \times N$ as defined in \eqref{Eq:code}, then ACCS between $C_{s}$ and $C_{t}$ is defined by
	\begin{equation}
		\mathcal{C}\left({C}_{s},{C}_{t}\right)(\tau)=\left\{\begin{array}{ll}
	\sum\limits_{i=1}^{M}\sum\limits_{j=1}^{N-\tau} c_{i,j}^s\left( c^{t}_{i,j+\tau}\right)^* , & \tau \ge 0\\
	\sum\limits_{i=1}^{M}\sum\limits_{j=1}^{N-\tau} c_{i,j-\tau}^s\left( c^{t}_{i,j}\right)^* , &  \tau<0.
		\end{array}\right. 
	\end{equation}
When ${C}_{s}={C}_{t}$, $\mathcal{C}(C_{s},C_{t})(\tau)$ is called AACS of ${C}_{s}$ and is denoted as $\mathcal{C}(C_{s})(\tau)$.
 \end{defn}
 \begin{defn}\label{def_SZCCS}
    Let $\mathbf{C}=\{C_1,C_2,\ldots,C_K\}$ be a set of $K$ codes, each having order $M \times N$ and it follows
	\begin{itemize}
	    \item[C1:] $\mathcal{C}(C_k)(\tau)=0$, for $|\tau|\in \mathcal{T}_1\bigcup\mathcal{T}_2$ and $1\le k\le K$,
     \item[C2:] $\mathcal{C}(C_{k_1},C_{k_2})(\tau)=0$, for $|\tau| \in \mathcal{T}_1\bigcup\mathcal{T}_2\bigcup\{0\}$, $k_1\ne k_2$ and $1\le k_1,k_2\le K$, 
	\end{itemize}
 where $\mathcal{T}_1=\{1,2,\ldots,Z\}$ and $\mathcal{T}_2=\{N-Z, N-Z+1,\ldots,N-1\}$
then set $\mathbf{C}$ is called a symmetrical $(K,M,N,Z)$-ZCCS. And when $K=\lfloor \frac{MN}{Z+1}\rfloor $, $\mathbf{C}$ is called an optimal symmetrical $(K,M,N,Z)$-ZCCS \cite{Zhou_sym_ZCCS_2023}.
\end{defn}
\begin{defn}
    Let $\mathbf{C}=\{C_1,C_2,\ldots,C_M\}$ be a set of $M$ codes of order $M \times N$ and it follows
    \begin{equation}
        \mathcal{C}(C_{k_1},C_{k_2})(\tau)=\left\{ \begin{array}{cc}
           MN  & k_1=k_2, \tau=0 \\
           0  & \text{otherwise.}
        \end{array}\right.
    \end{equation}
    Then $\mathbf{C}$ is called a $(M,N)$-CCC.
\end{defn}
\begin{defn}
    Let $\mathbf{C}_1, \mathbf{C}_2,\ldots, \mathbf{C}_K$ be collection of $(M,N)$-CCCs is referred as $(K,M,N,Z)$-MCCC if it follows
    \begin{equation}
        \mathcal{C}(C_{k_1},C_{k_2})(\tau)=0,
    \end{equation}
    where $C_{k_1}\in \mathbf{C}_i$, $C_{k_1}\in \mathbf{C}_j$ for $i\ne j$ and $|\tau|< Z$.
\end{defn}

Below, we present a lemma that is necessary in order to complete the proof.
\begin{lemma}[\cite{Rajen_direct_ZCCS}]\label{Lem:sum_zero}
For $M(\ge 2),s\in \mathbb{N}$, such that $M \nmid s$. Let $\zeta=\exp{\frac{2\pi \sqrt{-1}}{M}}$, then
\begin{equation}
    \sum_{i=1}^{M} \zeta^{s\cdot i}=0.
\end{equation}
\end{lemma}

\section{Proposed Construction}\label{Sec:PC}
\subsection{Construction using multiplication matrix}
The term "multiplication matrix" refers to a matrix in which the elements are determined by the product of subscripts.
\begin{construction}
    Let $s$ be any positive integer. Now, for a $2\le M \in \mathbb{N}$, we define a  matrix of size $M\times M$ as $\mathcal{M}(M,s,x)=\mathcal{M}$ where each entries of matrix is defined as
    \begin{equation}
        [\mathcal{M}]_{i,j}=\zeta^{x+s(i-1)(j-1)},
    \end{equation}
    for $i,j\in \{1,2,\ldots,M\}$, $x \in \mathbb{Z}$ and $\zeta=\frac{2\pi \sqrt{-1}}{M}$.
\end{construction}
\begin{thm}\label{Th:perfect_array}
    $\mathcal{M}(M,s,x)$ is a perfect array for $M,s,x \in \mathbb{Z}$, $M\ge 2$ and $M\nmid s$ ($M$ does not devides $s$).
\end{thm}
\begin{IEEEproof}
    Let $M,s,x \in \mathbb{Z}$, $M\ge 2$ and $\gcd(M,s)=1$. Then
    \begin{equation}\label{Eq:periodic_M}
        \mathcal{P}(\mathcal{M})(\tau_1,\tau_2)=\sum\limits_{i=1}^{M}\sum\limits_{j=1}^{M}\mathcal{M}_{i,j}(\mathcal{M}_{i+\tau_{1},j+\tau_{2}})^{*},
    \end{equation}
    where $\mathcal{M}_{i,j}=\zeta^{x+s(i-1)(j-1))}$ and subscript addition $i+\tau_1$ and $j+\tau_2$ are calculated over modulo $M$. When $(\tau_1,\tau_2)= (0,0)$ implies $\mathcal{M}_{i,j}(\mathcal{M}_{i+\tau_{1},j+\tau_{2}})^{*}=\mathcal{M}_{i,j}(\mathcal{M}_{i,j})*=1$ and $\mathcal{P}(\mathcal{M})(0,0)=M^2$. Let $(\tau_1,\tau_2)\ne (0,0)$
    \begin{equation}
    \begin{aligned}
        \mathcal{M}_{i,j}(\mathcal{M}_{i+\tau_{1},j+\tau_{2}})^{*}=&\zeta^{x+s(i-1)(j-1))-x-s(i+\tau_1-1)(j+\tau_2-1))}\\
        &=\zeta^{s(-i\tau_2-j\tau_1+\tau_1+\tau_2-\tau_1\tau_2)}.
    \end{aligned}
    \end{equation}
   Let $\Omega^1(\tau_1,\tau_2)=\tau_1+\tau_2-\tau_1\tau_2$. Now, we have
    \begin{equation}\label{Eq:shift_m}
    \begin{aligned} 
    \sum\limits_{i=1}^{M}\sum\limits_{j=1}^{M}\mathcal{M}_{i,j}(\mathcal{M}_{i+\tau_{1},j+\tau_{2}})^{*}=&\sum\limits_{i=1}^{M}\sum\limits_{j=1}^{M}\zeta^{-s(i\tau_2+j\tau_1-\Omega^1(\tau_1,\tau_2))}\\=&\zeta^{s\Omega^1(\tau_1,\tau_2)}\sum\limits_{i=1}^{M}\zeta^{-s i\tau_2}\sum\limits_{j=1}^{M}\zeta^{-s j\tau_1}
    \end{aligned}.
    \end{equation}
    From \eqref{Eq:shift_m} and \eqref{Eq:periodic_M}, we have
    \begin{equation}
        \begin{aligned}
         \mathcal{P}(\mathcal{M})(\tau_1,\tau_2)&=\zeta^{s\Omega^1(\tau_1,\tau_2)}\left(\sum\limits_{i=1}^{M}\zeta^{-s i\tau_2}\right)\left(\sum\limits_{j=1}^{M}\zeta^{-s j\tau_1}\right)\\
&=0, \text{from \textit{Lemma} \ref{Lem:sum_zero}.}
    \end{aligned} 
    \end{equation}
    This complete the proof.
\end{IEEEproof}
For $u\in \{0,1,\ldots,M-1\}$, $\mathcal{M}^u$ denotes the matrix obtained by repeating the cyclic shift with respect to the rows for $u$ times. 
\begin{thm}
    Let $M\ge 2$ be an arbitrary integer, $M\nmid s$ and $x\in\mathbb{Z}$. Then $\mathbf{C}=\{\mathcal{M}^u(M,s,x):u\in {0,1,\ldots M-1} \}$ is a $(M,M)$-CCC.
\end{thm}
\begin{IEEEproof}
For any $u\in \{0,1,\ldots,M-1\}$ each entries of $\mathcal{M}^u$ is defined as
\begin{equation}
    [\mathcal{M}^u]_{i,j}=[\mathcal{M}]_{i+u,j},
\end{equation}
where subscript addition $i+u$ is calculated modulo $M$ and $0$ is equivalent to $M$.

   Let $u_1,u_2,\tau \in \{0,1,\ldots,M-1\}$, then
   \begin{equation}
       \begin{aligned}
           [ \mathcal{M}^{u_1} ]_{i,j}[ \mathcal{M}^{u_2} ]^*_{i,j+\tau}=&[ \mathcal{M} ]_{i+u_1,j}[ \mathcal{M} ]^*_{i+u_2,j+\tau}\\
           =&\zeta^{s(i+u_1-1)(j-1)-s(i+u_2-1)(j+\tau-1)}\\
           =&\zeta^{s(j(u_1-u_2)-i\tau+\tau-\tau u_2+u_2)}\\
           =& \zeta^{sj(u_1-u_2)}\zeta^{-si\tau}\zeta^{\Omega^2(\tau,u_1,u_2)},
       \end{aligned}
   \end{equation}
   where $\Omega^2(\tau,u_1,u_2)=s(\tau-\tau u_2+u_2)$.
   \begin{equation}\label{Eq_pos_zero_shift}
      \begin{aligned}
           \mathcal{C}(\mathcal{M}^{u_1},\mathcal{M}^{u_2})(\tau)=&\sum\limits_{i=1}^{M}\sum\limits_{j=1}^{N-\tau}[ \mathcal{M}^{u_1} ]_{i,j}[ \mathcal{M}^{u_2} ]^*_{i,j+\tau}\\
=&\zeta^{\Omega^2}\sum\limits_{i=1}^{M}\zeta^{-si\tau}\sum\limits_{j=1}^{N-\tau}\zeta^{sj(u_1-u_2)}
      \end{aligned}
   \end{equation}
   Let $\tau=0$ and $u_1\ne u_2$, from \eqref{Eq_pos_zero_shift}, we have
   \begin{equation}
   \begin{aligned}
       \mathcal{C}(\mathcal{M}^{u_1},\mathcal{M}^{u_2})(0)=&\zeta^{\Omega^2}M\sum\limits_{j=1}^{M}\zeta^{sj(u_1-u_2)}\\=&0, \text{from \textit{Lemma} \ref{Lem:sum_zero}.}
   \end{aligned}
   \end{equation}
   Now, consider $\tau > 0$, from \eqref{Eq_pos_zero_shift}, we have
      \begin{equation}
   \begin{aligned}
       \mathcal{C}(\mathcal{M}^{u_1},\mathcal{M}^{u_2})(\tau)=&\zeta^{\Omega^2}\sum\limits_{j=1}^{N-\tau}\zeta^{sj(u_1-u_2)}\sum\limits_{i=1}^{M}\zeta^{-si\tau}\\
       =&0, \text{from \textit{Lemma} \ref{Lem:sum_zero}.}
   \end{aligned}
   \end{equation}
   Similarly, for $\tau < 0$, $\mathcal{C}(\mathcal{M}^{u_1},\mathcal{M}^{u_2})(\tau)=0$. This completes the proof.
\end{IEEEproof}
\subsection{Construction using mutually orthogonal sequences}
\begin{construction}
Let $C_1,C_2,\ldots,C_P$ be a set of $M\times N$ matrices and $\mathbf{b}=(b_1,b_2,\ldots,b_P)$ be sequence of length $P$, then we define 
\begin{equation}
    \mathcal{R}(C_1,C_2,\ldots,C_P;\mathbf{b})=[A_1||A_2||\cdots||A_P],
\end{equation}
where, $A_i=b_i \otimes C_i$ for $1\le i\le P$, $\otimes$ represents Kronecker product and $||$ represents concatenation of two matrices.
\end{construction}
Let $C_{i_1},C_{i_2},\ldots,C_{i_P},C_{k_1},C_{k_2},\ldots,C_{k_P}$ be a set of $M\times N$ matrices,  $\mathbf{b}^{j_1}=(b^{j_1}_1,b^{j_1}_2,\ldots,b^{j_1}_P)$ and $\mathbf{b}^{j_2}=(b^{j_2}_1,b^{j_2}_2,\ldots,b^{j_2}_P)$ be sequence of length $P$. Now, we construct two code as $B_1=\mathcal{R}(C_{i_1},C_{i_2},\ldots,C_{i_P};\mathbf{b}^{j_1})$ and $B_2=\mathcal{R}(C_{k_1},C_{k_2},\ldots,C_{k_P};\mathbf{b}^{j_2})$, then ACCS of $B_1$ and $B_2$ is obtained by
\begin{equation}\label{Eq:relation_cor_cons}
\begin{aligned}
     \mathcal{C}(B_1,B_2)(uN+v)=\sum_{\alpha=1}^{P-u} b^{j_1}_{\alpha} b^{{j_2}^*}_{\alpha+u} \mathcal{C} (C_{i_{\alpha}} , C_{k_{\alpha+u}})(v)\\
    +\sum_{\alpha=1}^{P-u-1}b^{j_1}_{\alpha}b^{{j_2}^*}_{\alpha+u+1}\left( \mathcal{C}(C_{i_{\alpha}},C_{k_{\alpha+u+1}})(v-N)\right) ,
\end{aligned}
\end{equation}
where  $0\le v <N$ and $-P\le u <P$.

\begin{thm}\label{Th:CCC_extension}
    Let $\mathbf{C}=\{C_1,C_2,\ldots,C_M\}$ be a $(M,N)$-CCC. Let $P$ and $S$ be two positive integers such that $PS=M$, for $1< P\le M$. Let $\mathbf{b}^1,\mathbf{b}^2,\ldots,\mathbf{b}^P$ be mutually orthogonal sequences of length $P$. Define 
    \begin{equation}
        B_{iP+j}=\mathcal{R}(C_{\pi(iP+1)},C_{\pi(iP+2)},\ldots,C_{\pi(iP+P)};\mathbf{b}^j),
    \end{equation}
where $0 \le i <  S$, $1\le j\le P$ and $\pi$ is permutation of $\{1,2,\ldots,M\}$. Then the obtained code set $\mathbf{B}=\{B_1,B_2,\ldots, B_M\}$ is a $(M,NP)$-CCC.
\end{thm}
\begin{IEEEproof}
    First, we prove that each code $B_k$ is a Golay complementary set (GCS), for $1\le k \le M$. Let $k=iP+j$, then AACF of $B_k$ is obtained by 
    \begin{equation}\label{Eq:AACF_B_k}
\begin{aligned}
    \mathcal{C}(B_k)(uN+v)=\sum_{\alpha=1}^{P-u}b^j_{\alpha}b^{j^*}_{\alpha+u}\mathcal{C}(C_{\pi(iP+\alpha)},C_{\pi(iP+\alpha+u)})(v)\\
    +\sum_{\alpha=1}^{P-u-1}b^j_{\alpha}b^{j^*}_{\alpha+u+1}\mathcal{C}(C_{\pi(iP+\alpha)},C_{\pi(iP+\alpha+u+1)})(v-N).
\end{aligned}
\end{equation}
Now, we discuss AACF for non-zero time shifts, i.e., excluding $u=v=0$.

Let $u=0$, implies $v\ne 0$ and $N-v \ne 0$. Putting $u=0$ in \eqref{Eq:AACF_B_k}, we have
\begin{equation}\label{Eq:AACF_B_k_u=0}
    \mathcal{C}(B_k)(0+v)=\sum_{\alpha=1}^{P}b^j_{\alpha}b^{j^*}_{\alpha}\mathcal{C}(C_{\pi(iP+\alpha)})(v).
\end{equation}
Each $C_{\pi(iP+\alpha)}$, for $\alpha \in \{1,2,\ldots,P\}$ is a GCS, hence $\mathcal{C}(C_{\pi(iP+\alpha)})(v)=0$, for $v\ne 0$, which makes \eqref{Eq:AACF_B_k_u=0} equals to zero.

Now, consider $u\ne 0$. As $C_{\pi(iP+\alpha)}$ is complementary mate with $C_{\pi(iP+\alpha+u)}$ (\& $C_{\pi(iP+\alpha+u+1)}$) for $1\le \alpha \le P-u\;(\& P-u-1)$, it implies $\mathcal{C}(C_{\pi(iP+\alpha)},C_{\pi(iP+\alpha+u)})(v)$ and $\mathcal{C}(C_{i,\pi(\alpha)},C_{i,\pi(\alpha+u+1-P)})(v-N)$ are zero in \eqref{Eq:AACF_B_k}. Therefore, each code $B_k$ is a GCS.

Let $1 \le k_1\ne k_2 \le M$, then $k_1=i_1P+j_1$ and $k_2=i_2P+j_2$ such that $(i_1,j_1)\ne(i_2,j_2)$. Then ACCS between $B_{k_1}$ and $B_{k_2}$ is as given in \eqref{Eq:ACCF_B_k_12}.

\begin{figure*}[t]
\begin{equation}\label{Eq:ACCF_B_k_12}
\begin{aligned}
    \mathcal{C}(B_{k_1},B_{k_2})(uN+v)=& \sum_{\alpha=1}^{P-u} b^{j_1}_{\alpha} b^{{j_2}^*}_{\alpha+u} \mathcal{C} (C_{\pi(i_1P+\alpha)} , C_{\pi(i_2P+\alpha+u)})(v)\\
    &+\sum_{\alpha=1}^{P-u-1}b^{j_1}_{\alpha}b^{{j_2}^*}_{\alpha+u+1}\left( \mathcal{C}(C_{\pi(i_1P+\alpha)},C_{\pi(i_2P+\alpha+u+1)})(v-N)\right) ,
\end{aligned}
\end{equation}
\end{figure*}
To complete the proof, we divided this case into two cases.
\begin{mycase}
    \case When $i_1=i_2$, i.e., $j_1\ne j_2$.
    \subcase For $u=v=0$,
    \begin{equation}
\begin{aligned}
    \rho(B_{k_1},B_{k_2})(0N+0)=&\sum_{\alpha=1}^{P}b^{j_1}_{\alpha}b^{{j_2}^*}_{\alpha}\rho(C_{\pi(i_1P+\alpha)})(0)\\
    =&MN \sum_{\alpha=1}^{P}b^{j_1}_{\alpha}b^{{j_2}^*}_{\alpha}=0.
\end{aligned}
\end{equation}
\subcase For $v=0$ but $u\ne 0$, $C_{\pi(i_1P+\alpha)}$ and $C_{\pi(i_2+\alpha+u)}$ are complementary mate for $1\le \alpha \le P$, which makes \eqref{Eq:ACCF_B_k_12} equals to zero.
\subcase $v\ne 0$. In this case $C_{\pi(i_1P+\alpha)}$ is either complementary mate of $C_{\pi(i_2P+\alpha+u)}$ or the same, for $1\le \alpha \le P$. In both cases $\rho(C_{\pi(i_1P+\alpha)},C_{\pi(i_2P+\alpha+u)})(v)=0$ for $v\ne 0$, which makes \eqref{Eq:ACCF_B_k_12} equals to zero.
\case When $i_1\ne i_2$. In this case, $C_{\pi(i_1P+\alpha)}$ and $C_{\pi(i_2P+\alpha+u)}$ are complementary mate, for $1\le \alpha \le P-u$, similarly, $C_{\pi(i_1P+\alpha)}$ and $C_{\pi(i_2P+\alpha+u+1)}$ are complementary mate, for $1\le \alpha < P-u-1$, which implies $\rho(C_{\pi(i_1P+\alpha)},C_{\pi(i_2P+\alpha+u)})(v)=0$ and $\rho(C_{\pi(i_1P+\alpha)},C_{\pi(i_2P+\alpha+u+1)})(N-v)=0$, which makes \eqref{Eq:ACCF_B_k_12} equals to zero.
\end{mycase}
Combining \textit{Case $1$} and \textit{Case $2$}, it is proved that $\rho(B_{k_1},B_{k_2})(\tau)=0$, for any value of $\tau$ and $k_1\ne k_2$. Therefore, $\mathbf{B}$ is a $(M,PN)$-CCC.
\end{IEEEproof}
 \begin{figure*}[!t]
 \begin{equation}\label{EQ:corr_B1_B2_symZCCS}
            \begin{aligned}
                  \mathcal{C}(B_{k_1M+i_1P+j_1},B_{k_2M+i_2P+j_2})(uN+v)=& \sum_{\alpha=1}^{P-u} b^{j_1}_{\alpha} b^{{j_2}^*}_{\alpha+u} \mathcal{C} (C_{\pi_{k_1}(i_1P+\alpha)} , C_{\pi_{k_2}(i_2P+\alpha+u)})(v)\\
    &+\sum_{\alpha=1}^{P-u-1}b^{j_1}_{\alpha}b^{{j_2}^*}_{\alpha+u+1}\left( \mathcal{C}(C_{\pi_{k_1}(i_1P+\alpha)},C_{\pi_{k_2}(i_2P+\alpha+u+1)})(v-N)\right) .
            \end{aligned}
        \end{equation}
    \end{figure*}
\begin{thm}\label{Th:multiCCC_SZCCS}
    Let $\mathbf{C}=\{C_1,C_2,\ldots,C_M\}$ be a $(M,N)$-CCC. Let $P$ and $S$ be two positive integers such that $PS=M$, for $1< P\le M$. Let $\mathbf{b}^1,\mathbf{b}^2,\ldots,\mathbf{b}^P$ be mutually orthogonal sequences of length $P$. Define 
    \begin{equation}
        B_{kM+iP+j}=\mathcal{R}(C_{\pi_k(iP+1)},C_{\pi_k(iP+2)},\ldots,C_{\pi_k(iP+P)};\mathbf{b}^j),
    \end{equation}
where $0 \le i <  S$, $1\le j\le P$, $0\le k < P-1$ and $\pi_k$ are permutation of $\{1,2,\ldots,M\}$. Each $\pi_k$ should be follow $\pi_{k_1}(i_1P+j)\ne \pi_{k_2}(i_2P+j)$ for $k_1\ne k_2$, $0 \le i_1,i_2 <S$ and $1\le j\le P$. Then, each $\mathbf{B}_1=\{B_1,B_2,\ldots, B_M\},$ $\mathbf{B}_2=\{B_{M+1},B_{M+2},\ldots, B_{2M}\},$ $\ldots$ $\mathbf{B}_P=\{B_{(P-1)M+1},B_{(P-1)M+2},\ldots, B_{PM}\}$ is a $(M,NP)$-CCC. Combining these sets $\mathbf{B}=\{\mathbf{B}_1\bigcup \mathbf{B}_2 \bigcup \cdots \bigcup \mathbf{B}_P\}$, is a MCCC also an optimal symmetrical $(PM,M,PN,N-1)$-ZCCS. 
\end{thm}
   
\begin{IEEEproof}
    From \textit{Theorem} \ref{Th:CCC_extension}, for a fixed $\pi_k$, corresponding $\mathbf{B}_k$ is a $(M,PN)$-CCC for $1\le k\le P$. Now, we show that codes from two different sets have an inter-set correlation zone. We observe the ACCF between $B_{k_1M+i_1P+j_1}$ and $B_{k_2M+i_2P+j_2}$, for $k_1\ne k_2$. From \eqref{Eq:relation_cor_cons} ACCF between $B_{k_1M+i_1P+j_1}$ and $B_{k_2M+i_2P+j_2}$ is as given in \eqref{EQ:corr_B1_B2_symZCCS}.

    The right hand side of \eqref{EQ:corr_B1_B2_symZCCS} is non-zero only for $v=0$, $\pi_{k_1}(i_1P+\alpha)=\pi_{k_2}(i_2P+\alpha+u)$ and $u\ne 0$, otherwise it is zero. And for $v=0$ and $u\ne 0$, the ACCF value maybe non-zero at only $|\tau|\in \{N,2N,\ldots,(P-1)N\}$. This completes the proof.
\end{IEEEproof}
\begin{example}
    Let $+$ represents $1$ and $-$ represents $-1$, now
 $        C_1=\begin{bmatrix}
       + + +\\
              + + - \\
              + + -\\
              - + -
        \end{bmatrix}, C_2=\begin{bmatrix}
             + - +\\
              + + - \\
               - - +\\
               + + +
        \end{bmatrix}, C_3=\begin{bmatrix}
            + - -\\ + + + \\ - + -\\+ - -
        \end{bmatrix}$ and $ C_4=\begin{bmatrix}
             + - -\\ + - + \\+ + + \\- + +
        \end{bmatrix}$ be a $(4,3)$-CCC \cite{Tao_SPL_ZCCS}. Here, $M=4$, $N=3$, assuming $P=2$, $\pi_1=(1,2,3,4)$, $\pi_2=(2,1,4,3)$ $\mathbf{b}^1=(--)$ and $\mathbf{b}^2=(+-)$. Now, from \textit{Theorem} \ref{Th:multiCCC_SZCCS}, we have $\mathbf{B}_1$ and $\mathbf{B}_2$ as two $(4,6)$-CCC given in Table \ref{Tab:ex_SZCCS}, which is also a $(2,4,6,3)$-MCCC and an optimal $(8,4,6,2)$-SZCCS. With the given parameter no CCC, MCCC and SZCCS are available.
        \begin{table}[!ht]
		\centering
			\caption{$(8,4,6,2)$-SZCCS}\label{Tab:ex_SZCCS}
		\begin{tabular}{ |c|c|c|c|c| }
			\hline
			& $B_1$ & $B_2$ & $B_3$ & $B_4$ \\
	\hline
        $\mathbf{B}_1$ &   $\begin{smallmatrix}
   -    -    -    -     +    -\\
    -    -     +    -    -     +\\
    -    -     +     +     +    -\\
     +    -     +    -    -    -

          \end{smallmatrix}$ & 
           $\begin{smallmatrix}
   +     +     +    -     +    -\\
     +     +    -    -    -     +\\
     +     +    -     +     +    -\\
    -     +    -    -    -    -
          \end{smallmatrix}$ &
          $\begin{smallmatrix}
   -     +     +    -     +     +\\
    -    -    -    -     +    -\\
     +    -     +    -    -    -\\
    -     +     +     +    -    -
          \end{smallmatrix}$ &
          $\begin{smallmatrix}
     +    -    -    -     +     +\\
     +     +     +    -     +    -\\
    -     +    -    -    -    -\\
     +    -    -     +    -    -

          \end{smallmatrix}$\\
         \hline
        & $B_5$ & $B_6$ & $B_7$ & $B_8$ \\
	\hline
  $\mathbf{B}_2$ &$\begin{smallmatrix}
      -     +    -    -    -    -\\
    -    -     +    -    -     +\\
     +     +    -    -    -     +\\
    -    -    -     +    -     +
          \end{smallmatrix}$ & 
           $\begin{smallmatrix}
      +    -     +    -    -    -\\
     +     +    -    -    -     +\\
    -    -     +    -    -     +\\
     +     +     +     +    -     +
          \end{smallmatrix}$ &
          $\begin{smallmatrix}
    -     +     +    -     +     +\\
    -     +    -    -    -    -\\
    -    -    -     +    -     +\\
     +    -    -    -     +     +
          \end{smallmatrix}$ &
          $\begin{smallmatrix}
     +    -    -    -     +     +\\
     +    -     +    -    -    -\\
     +     +     +     +    -     +\\
    -     +     +    -     +     +
          \end{smallmatrix}$\\
          \hline
         \end{tabular}
	\end{table}

\end{example}
\section{Comparison}\label{Sec:comparision}
In this section, we compare the proposed work with existing work.
\begin{table}[!ht]
    \centering
    \tiny
    \begin{tabular}{|c|c|c|}
    \hline
       Source  & Code set & Constraints \\
       \hline
        \cite{Tian_multiple_CCC} & $(2^v,2^{k+1},2^m,2^{m-v})$-MCCC & $0\le v \le k \le m-1$ \\
        \hline 
        \cite{Men_multiple_CCC} & $(p^v,p^n,p^m,p^{m-v})$-MCCC & $0\le v \le m$, $n\ge k+v$, \\
         & &$1\le k \le m-v$, $p$ is prime\\
        \hline
        Proposed & $(P,M,PN,N)$-MCCC & $(M,N)$-CCC available, $P|M$\\
         \hline
        \cite{Zhou_sym_ZCCS_2023} & $(2,2, 2^m+2^\delta,2^\delta-1)$-SZCCS &  $m\ge 3$, $1\le \delta \le m-1$\\
         & $(8,2,2^{m+1},2^{m-1}-1)$-SZCCS & \\
         \hline
         \cite{Praveen_sym_ZCCS} & $(p^{k+\delta},p^k,p^m,p^{m-\delta}-1)$-SZCCS & $m\ge3$, $1\le \delta \le m-1$\\
          & & $p$ is prime \\
        \hline
        Proposed & $(PM,M,PN,N-1)$-SZCCS & $(M,N)$-CCC available, $P|M$\\
        \hline
    \end{tabular}
    \caption{Comparison with existing MCCC and SZCCS}
    \label{tab:Comparision}
\end{table}

In the available work, the code size and set size appear to follow the power of prime, as the proposed construction accommodates any existing CCC, it results with additional parameters.
\section{Conclusion}\label{sec:conclusion}
In this paper, we introduce unique constructions for 2D perfect array and $(M,M)$-CCC utilising multiplication. In addition, new constructions have been developed for $(M,NP)$-CCC, $(P,M,PN,N)$ multiple CCC, and optimal $(PM,M,PN,N-1)$-SZCCS. These constructions utilise a $(M,N)$-CCC and mutually orthogonal sequences. The code set obtained offers more flexibility in comparison to the existing literature.
\IEEEtriggeratref{14}
 \bibliographystyle{IEEEtran}
 \bibliography{Rajen24}

\begin{thebibliography}{10}
\providecommand{\url}[1]{#1}
\csname url@samestyle\endcsname
\providecommand{\newblock}{\relax}
\providecommand{\bibinfo}[2]{#2}
\providecommand{\BIBentrySTDinterwordspacing}{\spaceskip=0pt\relax}
\providecommand{\BIBentryALTinterwordstretchfactor}{4}
\providecommand{\BIBentryALTinterwordspacing}{\spaceskip=\fontdimen2\font plus
\BIBentryALTinterwordstretchfactor\fontdimen3\font minus
  \fontdimen4\font\relax}
\providecommand{\BIBforeignlanguage}[2]{{%
\expandafter\ifx\csname l@#1\endcsname\relax
\typeout{** WARNING: IEEEtran.bst: No hyphenation pattern has been}%
\typeout{** loaded for the language `#1'. Using the pattern for}%
\typeout{** the default language instead.}%
\else
\language=\csname l@#1\endcsname
\fi
#2}}
\providecommand{\BIBdecl}{\relax}
\BIBdecl

\bibitem{Golomb_TIT_1982}
S.~Golomb and H.~Taylor, ``Two-dimensional synchronization patterns for minimum
  ambiguity,'' \emph{IEEE Trans. Inf. Theory}, vol.~28, no.~4, pp. 600--604,
  1982.

\bibitem{Hershey_EL_1983}
R.~Y. J.E.~Hershey, ``\BIBforeignlanguage{English}{Two-dimensional
  synchronisation},'' \emph{\BIBforeignlanguage{English}{Electronics Letters}},
  vol.~19, pp. 801--803(2), September 1983.

\bibitem{Golay_71}
M.~J.~E. Golay, ``Point arrays having compact, nonredundant autocorrelations,''
  \emph{J. Opt. Soc. Am.}, vol.~61, no.~2, pp. 272--273, Feb 1971.

\bibitem{Fan1996SEQUENCE}
P.~Fan and M.~Darnell, \emph{Sequence design for communications
  applications}.\hskip 1em plus 0.5em minus 0.4em\relax John Wiley \& Sons
  Inc., 1996.

\bibitem{Weathers_1983}
G.~Weathers and E.~Holliday, ``Group-complementary array coding for radar
  clutter rejection,'' \emph{IEEE Trans. Aerospace and Electronic Systems},
  vol. AES-19, no.~3, pp. 369--379, 1983.

\bibitem{Trucs}
M.~Turcs{\'a}ny and P.~Farka{\v{s}}, ``New {2D-MC-DS-SS-CDMA} techniques based
  on two-dimensional orthogonal complete complementary codes,'' in
  \emph{Multi-Carrier Spread-Spectrum}, K.~Fazel and S.~Kaiser, Eds.\hskip 1em
  plus 0.5em minus 0.4em\relax Dordrecht: Springer Netherlands, 2004, pp.
  49--56.

\bibitem{sarkar2024construction}
\BIBentryALTinterwordspacing
P.~Sarkar, C.~Li, S.~Majhi, and Z.~Liu, ``Construction of complete
  complementary codes over small alphabet,'' 2024. [Online]. Available:
  \url{https://arxiv.org/abs/2102.10517}
\BIBentrySTDinterwordspacing

\bibitem{Praveen_CCC}
P.~Kumar, S.~Majhi, and S.~Paul, ``A direct construction of {G}olay
  complementary pairs and binary complete complementary codes of length
  non-power of two,'' \emph{IEEE Trans. Commun.}, vol.~71, no.~3, pp.
  1352--1363, 2023.

\bibitem{Tao_SPL_ZCCS}
Y.~Tao, A.~Avik, Ranjan, W.~Yanyan, and Y.~Yang, ``New class of optimal
  {Z}-complementary code sets,'' \emph{IEEE Signal Process. Lett.}, vol.~29,
  pp. 1477--1481, 2022.

\bibitem{Fan}
P.~Fan, W.~Yuan, and Y.~Tu, ``Z-complementary binary sequences,'' \emph{IEEE
  Signal Process. Lett.}, vol.~14, no.~8, pp. 509--512, 2007.

\bibitem{Feng_ZCCS}
L.~Feng, P.~Fan, X.~Tang, and K.-k. Loo, ``Generalized pairwise
  {Z}-complementary codes,'' \emph{IEEE Signal Process. Lett.}, vol.~15, pp.
  377--380, 2008.

\bibitem{Rajen}
\BIBentryALTinterwordspacing
R.~Kumar, S.~K. Jha, P.~K. Srivastava, and S.~Majhi, ``Construction of type-{II
  ZCCS} for the {MC-CDMA} system with low {PMEPR},'' \emph{Digital Signal
  Processing}, vol. 151, p. 104570, 2024. [Online]. Available:
  \url{https://www.sciencedirect.com/science/article/pii/S1051200424001957}
\BIBentrySTDinterwordspacing

\bibitem{Zhou_sym_ZCCS_2023}
Y.~Zhou, Z.~Zhou, Z.~Liu, Y.~Yang, P.~Yang, and P.~Fan, ``Symmetrical
  {Z}-complementary code sets for optimal training in generalized spatial
  modulation,'' \emph{Signal Processing}, vol. 208, p. 108990, 2023.

\bibitem{Tian_multiple_CCC}
L.~Tian, Y.~Li, and C.~Xu, ``Multiple complete complementary codes with
  inter-set zero cross-correlation zone,'' \emph{IEEE Trans. Commun.}, vol.~68,
  no.~3, pp. 1925--1936, 2020.

\bibitem{Rajen_direct_ZCCS}
\BIBentryALTinterwordspacing
R.~Kumar, P.~K. Srivastava, and S.~Majhi, ``A construction of arbitrarily large
  type-{II $Z$} complementary code set,'' 2024. [Online]. Available:
  \url{https://arxiv.org/abs/2305.01290}
\BIBentrySTDinterwordspacing

\bibitem{Men_multiple_CCC}
X.~Men and Y.~Li, ``New construction of multiple complete complementary codes
  with inter-set zero cross-correlation zone,'' \emph{IEEE Signal Processing
  Letters}, vol.~29, pp. 1958--1962, 2022.

\bibitem{Praveen_sym_ZCCS}
P.~Kumar, S.~Majhi, and S.~Paul, ``A direct construction of optimal symmetrical
  {Z}-complementary code sets of prime power lengths,'' in \emph{IEEE
  International Symposium on Information Theory (ISIT)}, 2023, pp. 1283--1287.

\end{thebibliography}
\end{document}